\documentclass[prl,twocolumn,showpacs]{revtex4-1}
\usepackage{graphicx}
\usepackage{amssymb,amsmath}
\usepackage{bm}
\usepackage[applemac]{inputenc}
\usepackage{pdfpages}

\begin{document}    
  
\title{ac Josephson effect in finite-length nanowire junctions with Majorana modes}

\author{Pablo San-Jose$^1$, Elsa Prada$^2$, Ramón Aguado$^2$}
\affiliation{$^1$Instituto de Estructura de la Materia (IEM-CSIC), Serrano 123, 28006 Madrid, Spain\\$^2$Instituto de Ciencia de Materiales de Madrid (ICMM-CSIC), Cantoblanco, 28049 Madrid, Spain}

\date{\today} 


\begin{abstract}

It has been predicted that superconducting junctions made with topological nanowires hosting Majorana bound states (MBS) exhibit an anomalous $4\pi$-periodic Josephson effect. Finding an experimental setup with these unconventional properties poses, however, a serious challenge:  for finite-length wires, the equilibrium supercurrents are always $2\pi$-periodic as anticrossings of states with the same fermionic parity are possible. We show, however, that the anomaly survives in the transient regime of the ac Josephson effect. Transients are moreover protected against decay by quasiparticle poisoning as a consequence of the quantum Zeno effect, which fixes the parity of Majorana qubits. The resulting long-lived ac Josephson transients may be effectively used to detect MBS.
\end{abstract}
 

\maketitle  
%Majorana fermions, a real solution of the Dirac equation discovered by Ettore Majorana in 1937, may appear in condensed matter as emergent low-energy excitations \cite{Wilczek:NP09,Franz:P10,Hughes:P11,Service:S11} with Non-Abelian statistics \cite{Ivanov:PRL01,Stern:N10}. Early examples of materials predicted to host such excitations include exotic superconductors with p-wave pairing~\cite{Volovik:JL99,Read:PRB00}.  
Recently, it has been argued that Majorana bound states (MBS) should appear in topological insulators~\cite{Fu:PRL08} and semiconductors with strong spin-orbit (SO) coupling~\cite{Sau:PRL10,Alicea:PRB10,Lutchyn:PRL10,Oreg:PRL10} which, in proximity to s-wave superconductors, may behave as topological superconductors (TS). 
MBS in these TS can be understood as Bogoliubov-De Gennes (BdG) quasiparticles appearing inside the superconducting gap, exactly at zero energy (for a review see Ref. \onlinecite{Beenakker:11}). %MBS come in pairs and are located at the edges of the nanowire (NW) or wherever the system interfaces with a non-topological phase \cite{Alicea:NP11}.
The TS phase is tunable, which has spurred a great deal of experimental activity towards detecting MBS. %in hybrid superconductor-semiconductor systems 
Tunneling through such TS is expected to show a zero-bias anomaly signaling the presence of MBS \cite{Bolech:PRL07,Nilsson:PRL08,Law:PRL09,Flensberg:PRB10,Wimmer:NJOP11,Gibertini:12,Prada:12}. However, such anomaly, which has been recently observed \cite{Delft-exp}, only proves a necessary condition. A a more stringent test can be established by measuring the Josephson effect through a junction between two TS nanowires. %  (for alternative schemes see Refs. \cite{Bolech:PRL07,Nilsson:PRL08,Law:PRL09,Flensberg:PRB10,  Fu:PRL10,Wimmer:NJOP11,Zazunov:PRB11}). 
Kitaev predicted \cite{Kitaev:P01} that such Josephson effect has an anomalous $4\pi$-periodicity in the superconducting (SC) phase difference $\phi\equiv\phi_1-\phi_2$ between the two wires. This \emph{fractional} Josephson current
 $I_J\sim \sin (\frac{\phi}{2})$ %may be interpreted as tunneling of half a Cooper pair through the zero-energy MBS inside the superconducting gap and is ubiquitous in SC junctions with MBS 
\cite{Lutchyn:PRL10,Oreg:PRL10,Kitaev:P01,Kwon:EPJB03,Fu:PRB09,Ioselevich:PRL11,Law:PRB11,Jiang:11} %. The $4\pi$-periodicity 
can be understood in terms of fermion parity (FP):
%if one considers an infinitely-long NW, the relevant physics can be understood as two MBS fusing across the junction and forming a fermion state which can be either empty (even parity) or full (odd parity). 
if FP is 
\emph{preserved}, there is a protected crossing of Majorana states at $\phi=\pi$ \footnote{For the sake of the argument, we use a fixed value $\phi=\pi$ for the zero-energy crossing although its position is not universal, see, e. g. \cite{Lutchyn:PRL10}} with perfect population inversion, namely the system \emph{cannot} remain in the ground state as $\phi$ evolves from 0 to $2\pi$ adiabatically, unlike for standard Andreev bound states (ABS) \cite{resonantJosephson}.
%the reason being that the ground state parity changes when $\phi\rightarrow\phi+2\pi$. 
 % due to the orthogonality of the two Majorana states. 
%Finite length effects, however, render the Josephson effect $2\pi$-periodic~\cite{Pikulin:JL11}: 
However, any finite length of the two TS regions gives rise to two additional MBS at the wire ends (which are connected to topologically trivial regions), allowing for the hybridization of two states of the same FP. This results in residual splittings %(\emph{anticrossings}) 
at $\phi=\pi$ which, %Under slow modulation of $\phi$, 
despite being exponentially small, %, $\sim e^{-L_{S'}/\xi}$, with $\xi$ being the superconducting coherence lenght) 
destroy the fractional effect as the system remains in the ground state for all $\phi$ ~\cite{Pikulin:JL11}.
%As it turns out, the 
%This splitting arises by the hybridization of the two MBS in the junction with two additional MBS that develop at the opposite end of each wire, which are linked to the (topologically trivial) external circuit.

%%%%%%%%%%%%%%%%%%%%%%%%%%%%%%%%%%%%%%%%%%%%%%%%%%
\begin{figure}
  \centering
  \includegraphics[width=0.38\textwidth]{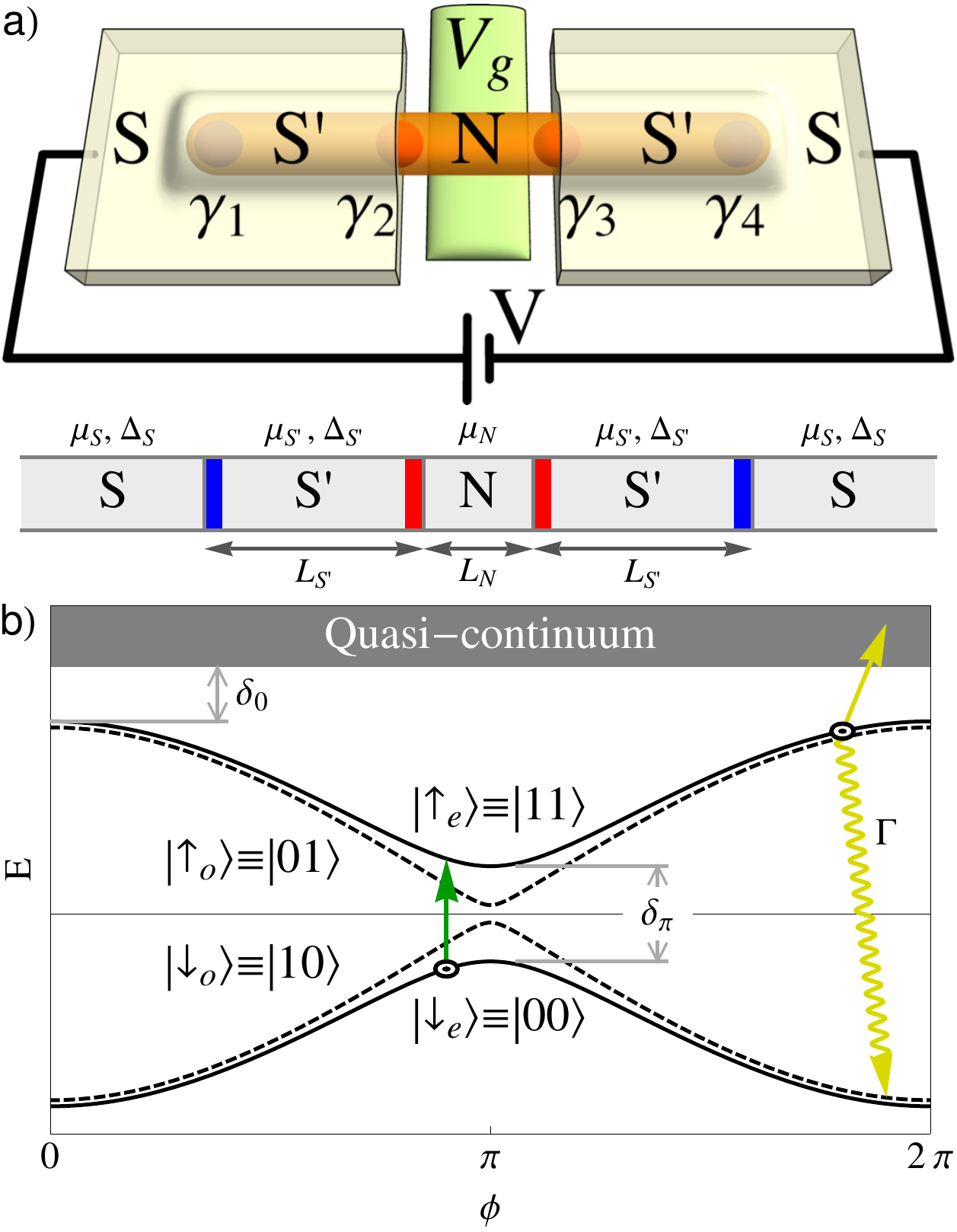}
\caption{(Color online) (a) Schematics: A nanowire of length $L=L_{S'}+L_N+L_{S'}$ in contact with two s-wave superconductors (with gap $\Delta_S$), develops a proximity-induced SC gap $\Delta_{S'}<\Delta_S$ and four Majorana modes $\gamma_{1,2,3,4}$. The transparency of the junction can be controlled by a gate voltage $V_g$. (b) Energy of the four lowest many-body states in the TS phase, as a function of SC phase difference $\phi$ across $N$. They correspond to even (solid) and odd (dashed) fillings of the two lowest ABS, see Fig. \ref{fig:2}b. Note the avoided crossing of size $\delta_\pi$ at $\phi=\pi$, and the detachment gap $\delta_0$ at $\phi=0$.
%, that is due to the  hybridization of inner and outer Majorana modes (red and blue in (a), respectively). 
%Driving $\phi$ with a bias voltage $V$, induces parity conserving transitions around $\phi=2\pi(n+1/2)$ (integer $n$), and parity breaking dissipative processes around $\phi=2\pi n$, involving transitions into the quasi-continuum across the detachment gap $\delta_0$, which may be controlled through $V_g$.
}
  \label{fig:1}
\end{figure}
%%%%%%%%%%%%%%%%%%%%%%%%%%%%%%%%%%%%%%%%%%%%%%%%%%

The $4\pi$-periodicity can be restored in two ways. On the one hand, one may employ a fully topological circuit \cite{Heck:PRB11}, which is free of additional MBS. In other words, by employing an S'NS' geometry (where S' is a TS and N is a normal, non-topological region), as opposed to an SS'NS'S junction (where S is a topologically trivial superconductor).  
Unfortunately, this is presently a difficult experimental challenge using nanowires. Alternatively, the $4\pi$ Josephson effect may be recovered by biasing the junction and thereby sweeping $\phi$ fast enough through the anticrossing such that Landau-Zener (LZ) processes \cite{Wittig:JPCB05} induce non-adiabatic transitions between states of the same FP.  %An obvious choice to accomplish this is to apply a dc bias voltage $V_{dc}$ which establishes a time-dependent phase drop $\phi=2eV_{dc}t/\hbar$ across the junction. 
By the same token, however, transitions to the continuum of states above the gap, and hence parity mixing processes, will be unavoidable \cite{Badiane:PRL11}.
%Such processes are expected to once more spoil the fractional Josephson effect by parity mixing (quasiparticle poisoning) \cite{Badiane:PRL11}.

The above hindrances pose relevant questions. Is the Josephson effect a useful tool to detect MBS in realistic TS wires? Do finite TS length and quasiparticle poisoning inevitably destroy the fractional periodicity? We here prove that the ac currents in the TS phase contain anomalously long-lived $4\pi$-periodic transients which are tunable through both bias and gate voltages. Thus, the Josephson effect may still be exploited to provide unequivocal proof of the existence of MBS in finite-length TS. %The dynamical regimes of the TS junction are summarized in a diagram of the ac current %as a function of the Josephson frequency and the transmission of the normal part, 
%with large regions of $4\pi$-periodicity which are tunable through both bias and gate voltages. % with a superimposed beating envelope owing to incomplete LZ transitions. These regions are tunable through both bias and gate voltages.
%, albeit through the spectral analysis of Josephson current \emph{transients}, which develop $4\pi$-periodic components. We further show that such transients may be anomalously long-lived in the TS phase. A notable result of our study is that quasiparticle escape towards the contacts may actually help to fix fermion parity, delaying the decay of the anomalous transient regime, and facilitating its detection. %Our idea builds upon the concept that, in some cases, dissipation may be helpful in boosting or even inducing new quantum effects.  In particular, we use 
Interestingly, the duration of the long $4\pi$-periodic transients may be increased by the well-known, though somewhat counter-intuitive, \emph{quantum Zeno effect}, whereby a strong coupling to a decohering environment helps to confine the dynamics of a quantum system into a desired sector of the Hilbert space \cite{Facchi:PRL02}.% (quantum Zeno subspace \cite{Facchi:PRL02}). 

\begin{figure}
\centering
 \includegraphics[width=0.48\textwidth]{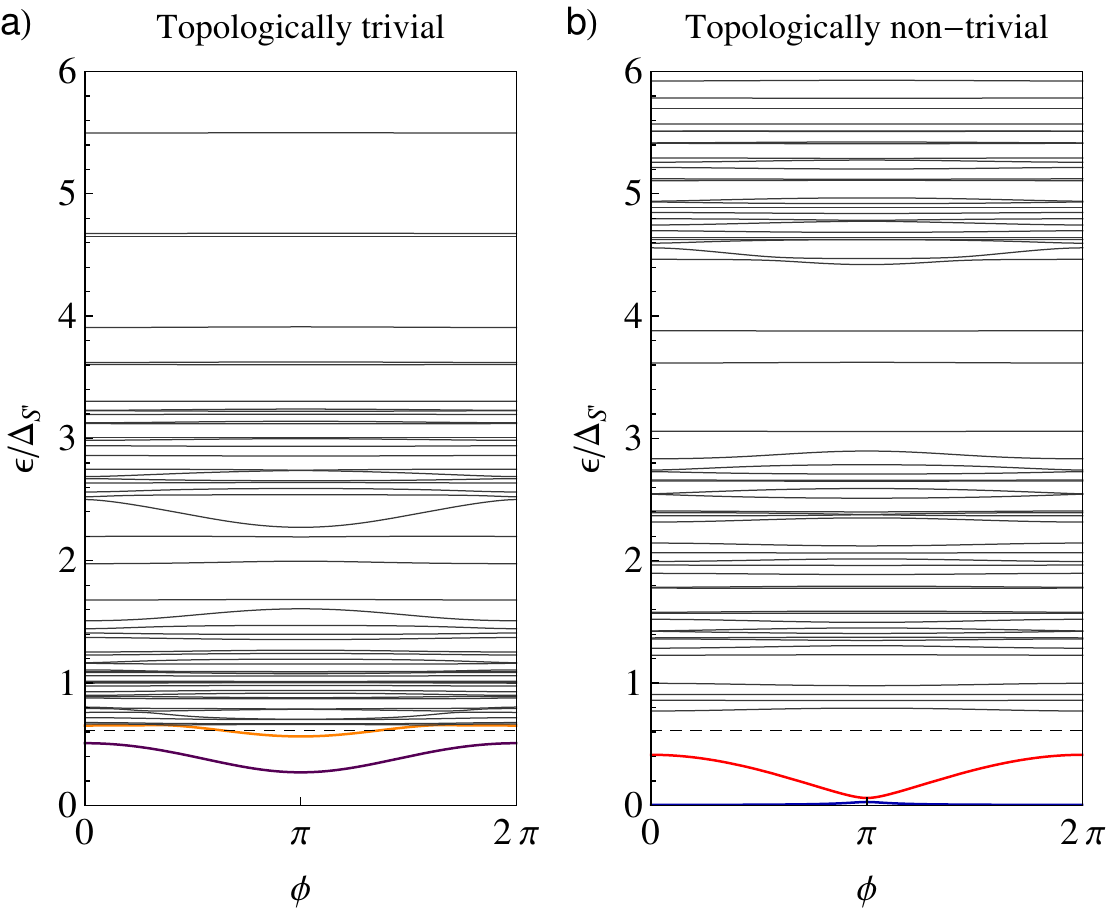}
  \caption{(Color online) Andreev bound states (ABS) for a SS'NS'S junction with normal conductance $G=0.72G_0$ (with $G_0=2e^2/h$), an induced gap $\Delta_{S'}=218 \mu\mathrm{eV}$, and wire's length of $4 \mu$m (where $l_{so}=216$nm for a InSb wire). In the topologically trivial phase (a), $\mathcal{B}=0.36 \mathrm{ meV}=0.5\mathcal{B}_c$, whereas in the non-trivial phase (b), $\mathcal{B}=1.1\mathrm{ meV}=1.5 \mathcal{B}_c$. The dashed line denotes the wire effective gap $\Delta_{\rm{eff}}=130\mu\mathrm{eV}$ that separates localized ABS from the quasicontinuum. 
  %In the topological case (b), $\Delta_{\rm{eff}}$ is defined as the smallest of the two gaps $\Delta_1\equiv |\mathcal{B}-\mathcal{B}_c|$ and $\Delta_2\equiv\Delta_{S'}\sqrt{\frac{2(1+\tilde{\mu}+\sqrt{1+2\tilde{\mu}+\tilde{\mathcal{B}}^2})}{\tilde{\mathcal{B}}^2+2(1+\tilde{\mu}+\sqrt{1+2\tilde{\mu}+\tilde{\mathcal{B}}^2})}}+\mathcal{O}(\Delta_{S'}^2)$, with tilde quantities denoting energies in units of the SO energy, $E_{so}\equiv\hbar^2/ml_{so}^2$, with $l_{so}\equiv\hbar^2/m^*\alpha$ the SO length given in terms of the wire's effective mass $m^*$ and SO coupling $\alpha$.
%Panel (c) shows the limited range of excitation probabilities of a fermion (starting in Andreev level $d^\dagger_2|\Omega\rangle$) into the lower quasi-continuum levels after a single pass across $\phi=2\pi$ (the dashed curve is the two-level LZ prediction for the fastest rate).
} 
 \label{fig:2}
\end{figure}

\paragraph{Low-energy description.--}% of finite length topological superconductors in terms of Majorana logical qubits}
Consider a semiconducting one-dimensional wire with chemical potential $\mu_{S'}$, SO coupling $\alpha$, and Zeeman splitting $\mathcal{B}$ (given by $\mathcal{B}=g\mu_BB/2$, where $B$ is an in-plane magnetic field, $\mu_B$ is the Bohr magneton and $g$ is the wire g-factor). When the wire is proximity coupled to an s-wave superconductor of gap $\Delta_S$, a SC pairing term $\Delta_{S'}<\Delta_S$ may be induced. The wire is then arranged into a Josephson device as illustrated in Fig. \ref{fig:1}a. The resulting hybrid system SS'NS'S can be driven into a TS phase where two MBS are formed at the ends of each $S'$ region, denoted by $\gamma_{1,2}$ and $\gamma_{3,4}$, when $\mathcal{B}>\mathcal{B}_c\equiv\sqrt{\mu_{S'}^2+\Delta_{S'}^2}$~\cite{Lutchyn:PRL10,Oreg:PRL10}. %(neglecting interactions \cite{Gangadharaiah:PRL11,Sela:PRB11,Stoudenmire:PRB11}). 
The low-energy Majorana sector appears for energies $\varepsilon<\Delta_{\rm{eff}}$, where $\Delta_{\rm{eff}}$ is the effective SC gap (Fig. \ref{fig:2}b). 

Isolated MBS $\gamma_i$ are zero energy superpositions of a particle and a hole. %, e.g. $\gamma_1=(c^++c)/\sqrt{2}$ or $\gamma_2=i(c^+-c)/\sqrt{2}$. %(MBS are their own antiparticle $\gamma_i=\gamma_i^+$). 
A pair of Majoranas $\gamma_{1,2}$ may be fused into a Dirac fermion $c^\dagger=(\gamma_1-i\gamma_2)/\sqrt{2}$, such that the operator $2c^\dagger c-1=2i\gamma_1\gamma_2$ with eigenvalues -1 and 1 in states $|0\rangle$ and 
$|1\rangle=c^\dagger|0\rangle$ defines FP \cite{MajoranaQubits}.
%\label{qubit}
%\end{eqnarray}
Thus, a spatial overlap of two MBS hybridizes them into eigenstates of opposite FP.
In a TS wire of length $L_{S'}$, the decay distance of the two MBS pinned at the wire ends is the effective coherence length $\xi_{\rm{eff}}=\hbar v_F/\pi \Delta_{\rm{eff}}$. Their overlap will induce a splitting $\sim\pm\Delta_{\rm{eff}}e^{-L_{S'}/\xi_{\rm{eff}}}$. Similarly, two MBS at either side of a Josephson junction with phase difference $\phi$ and transparency $T$ will hybridize into even/odd fermion states with energies $\sim\pm\Delta_{\mathrm{eff}}\sqrt{T}\cos\phi/2$, \cite{Kwon:EPJB03,Fu:PRB09,Badiane:PRL11}. % which yields a parity inversion of the ground state as $\phi\to \phi+2\pi$.
In the setup of Fig. \ref{fig:1}, \emph{four} MBS (two `inner' $\gamma_{2,3}$ and two `outer' $\gamma_{1,4}$) hybridize both through the Josephson junction (region $N$) and the finite length $S'$ regions. The resulting eigenstates are empty and filled states of two Dirac fermions $d^\dagger_{1,2}(\phi)$, constructed as two $\phi$-dependent (orthogonal) superpositions of the two fermions $c_{in}^\dagger=(\gamma_2+i\gamma_3)/\sqrt{2}$ and $c_{out}^\dagger=(\gamma_1+i\gamma_4)/\sqrt{2}$, which are themselves obtained from the fusion of the inner and outer MBS, respectively. We denote eigenenergies by $E_{n_1n_2}$ and eigenstates as $|n_1 n_2\rangle$, where $n_1, n_2=0,1$ are the occupations of fermions $d_1^\dagger$ and $d_2^\dagger$. Two of them have even \emph{total parity}, $|\downarrow_{e}\rangle\equiv|00\rangle$, $|\uparrow_{e}\rangle\equiv|11\rangle=d_2^\dagger d_1^\dagger|00\rangle$, and the other two are odd, $|\downarrow_{o}\rangle\equiv|10\rangle=d_1^\dagger|00\rangle$, $|\uparrow_{o}\rangle\equiv|01\rangle= d_2^\dagger|00\rangle$ \cite{logicalQB}. $E_{n_1n_2}(\phi)$ anti-cross at $\phi=\pi$ within same-parity sectors (Fig. \ref{fig:1}b) and, hence, the supercurrents are $2\pi$-periodic. 

A $4\pi$-periodic Josephson effect can, nevertheless, be recovered by inducing LZ transitions with a voltage bias, such as $|\downarrow_{e}\rangle\rightarrow|\uparrow_{e}\rangle$ (green arrow in Fig. 1b) \cite{Pikulin-Nazarov} . To describe the response of a realistic biased junction, however, an extension of the simplified Majorana model above is required. Indeed, non-adiabatic driving may also induce inelastic transitions into delocalized states above the TS gap (yellow arrow in Fig. \ref{fig:1}b). These latter transitions induce an effective parity-mixing rate (yellow wiggly arrow) which couples even and odd sectors (quasiparticle poisoning). A proper description of such dynamics involves a calculation of \emph{all} the Andreev levels (both below and above $\Delta_{\rm{eff}}$) coupled to the continuum (above $\Delta_S$) of the junction, which we develop in what follows. 

%%%%%%%%%%%%%%%%%%%%%%%%%%%%%%%%%%%%%%%%%%%%%%%%%%
%%%%%%%%%%%%%%%%%%%%%%%%%%%%%%%%%%%%%%%%%%%%%%%%%%
 
\paragraph{Andreev levels.--}The full spectrum of single-particle eigenstates may be obtained by diagonalizing the BdG equations for the geometry in Fig. 1a \cite{SI}. We obtain $H_{BdG}=\frac{1}{2}\sum_n\left(d^\dagger_n d_n-d_n d^\dagger_n\right)\varepsilon_n$, with eigenenergies $\varepsilon_{n}(\phi)$ plotted in Fig. \ref{fig:2}a,b for a representative junction. The corresponding single particle excitations, called Andreev bound states (ABS), are defined as $|n(\phi)\rangle=d^\dagger_n(\phi)|\Omega(\phi)\rangle$, with $|\Omega(\phi)\rangle$ denoting the ground state. In the short junction limit, $L_{N}\ll\xi_{\rm{eff}}$, only two ABS lie below the effective gap $\Delta_{\rm{eff}}$. Upon crossing into the TS phase, these two levels, initially localized around the $N$ region in the non-topological phase (purple and orange curves in Fig. \ref{fig:2}a), reconnect into the two $d^\dagger_{1,2}(\phi)$ Majorana branches.
The lowest Majorana branch (blue curve in Fig. \ref{fig:2}b, level $d^\dagger_1$) has a weak $\phi$ energy dependence, and is dominated by the outer MBS ($c_{out}^\dagger$ fermion, $\varepsilon_1\approx \Delta_{\mathrm{eff}}e^{-L_{S'}/\xi_{S'}}$), whereas the one at higher energy (red curve, $d^\dagger_2$) results mostly from the fusion of the inner MBS ($c_{in}^\dagger$ fermion, $\varepsilon_2\approx \Delta_{\mathrm{eff}}\sqrt{T}\cos\phi/2$).  

Due to the finite length $L_{S'}>\xi_{S'}$, a dense set (quasicontinuum) of ABS delocalized across the TS wire appears within a large energy window $\Delta_{S}>\varepsilon>\Delta_{\rm{eff}}$ above the low-energy sector. This quasicontinuum (which approaches a continuum as $L_{S'}\rightarrow\infty$), separates the Majorana branches from the true continuum of the problem at $\varepsilon>\Delta_{S}$. Importantly, for finite transparency junctions, the Majorana levels detach from the quasicontinuum, as opposed to the topologically trivial case or a standard SNS junction \cite{Averin:PRL95} where the ABS  touch the continuum at $\phi=0,2\pi$. We have verified that the detachment, denoted by $\delta_0$ in Fig. \ref{fig:1}b, increases as the transmission $T$ of the normal region becomes smaller, $\delta_0/\Delta_{\mathrm{eff}}\approx 1-\sqrt{T}$, in agreement with simpler models \cite{Kwon:EPJB03,Fu:PRB09,Badiane:PRL11}.  %Parity mixing in the low energy Majorana sector by quasiparticle escape into the quasi-continuum is, therefore, mediated by coherent transitions across the gap $\delta_0$, which profoundly affects the parity dynamics, as shown in what follows.
%%%%%%%%%%%%%%%%%%%%%%%%%%%%%%%%%%%%%%%%%%%%%%%%%%
\begin{figure}
 \centering
 \includegraphics[width=0.4\textwidth]{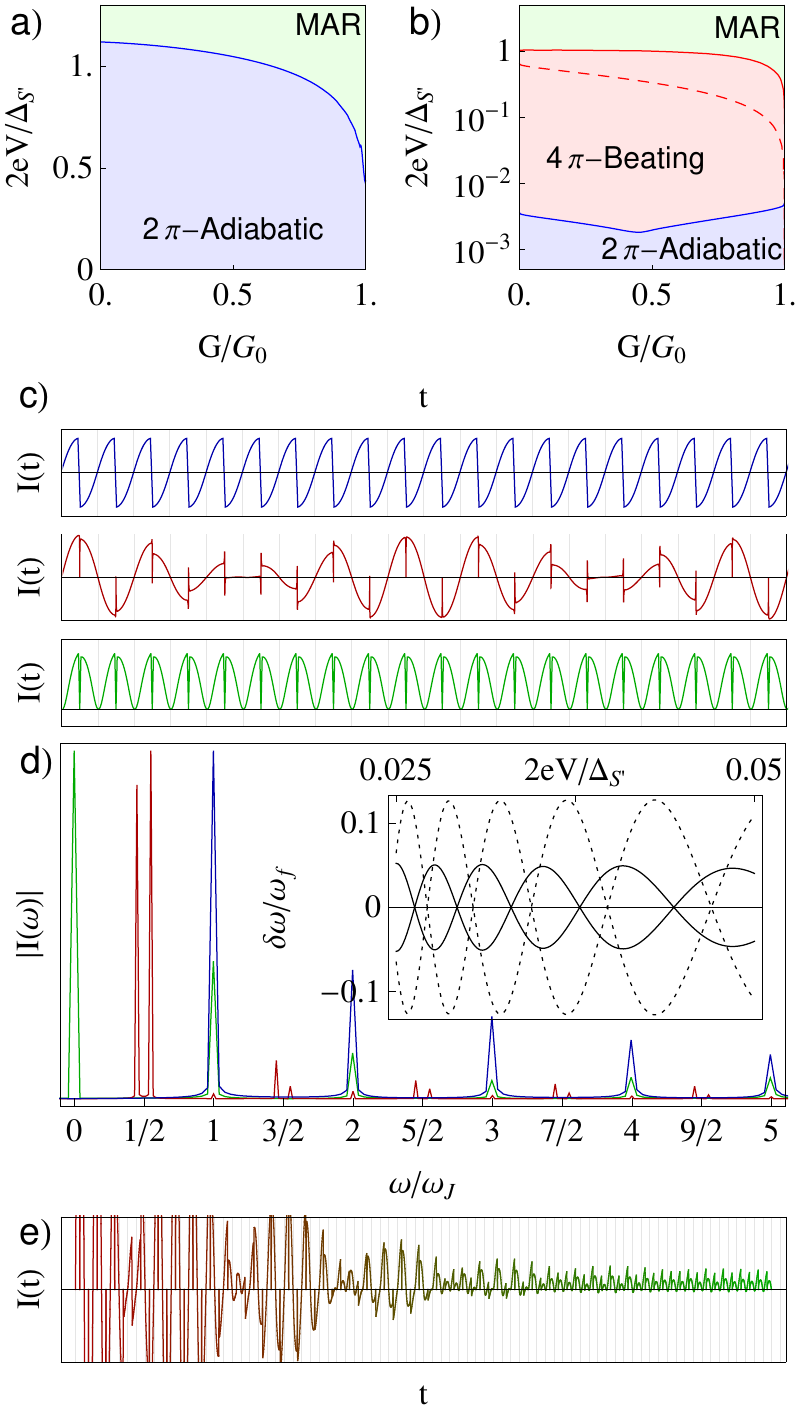}
 \caption{(Color online) Top: diagrams of the different dynamical regimes of the ac Josephson current as a function of junction bias $V$ and dimensionless conductance $G/G_0$, for  (a) the topologically trivial  and (b) non-trivial phases of a $8.7\mu$m long InSb wire similar to that of Fig. \ref{fig:2}. Typical current profiles are presented in (c). Conventional adiabatic (blue, top) and multiple Andreev reflection (MAR) (green, bottom) regimes are both $2\pi$-periodic. A wide region of beating $4\pi$-periodic ac Josephson current (red, middle) appears in the non-adiabatic TS phase, and is absent in the trivial case. Each regime has a distinct frequency transform (d). %The fractional Josephson is revealed as a long-lived, transient regime that exhibits a superimposed beating pattern, evident as a split $\omega_J/2$ peak in the frequency spectrum. 
%Panel  presents a typical dissipative decay from the $4\pi$ regime into the stationary MAR regime, which develops a dc current component (peak at $\omega=0$). 
The inset shows the beat splitting $\pm\delta\omega$ of the anomalous peak at $\omega_J/2$ as a function of the bias $V$, both for the even (solid) and the odd sectors (dashed).
Panel (e)  presents a typical dissipative decay from the $4\pi$-regime into the stationary MAR regime, which develops a dc current component (peak at $\omega=0$).
}
 \label{fig:3}
\end{figure}
\paragraph{ac Josephson effect.--}
We now study the Josephson current across the junction when biased with a voltage $V$. The bias makes the phase difference $\phi$ time dependent, $\phi(t)=2eVt/\hbar=\omega_J t$, where $\omega_J$ is the Josephson frequency. 
%Since the levels $\varepsilon_{m\geq 3}$ above $\Delta_{\mathrm{eff}}$ are almost $\phi$ independent, 
We calculate the Josephson current through the biased junction in terms of a \emph{reduced} density matrix in the Majorana sector. Notably, its evolution contains parity-mixing terms as a result of the Andreev levels above $\Delta_{\mathrm{eff}}$ that have been traced out \cite{SI}.
%namely the $\tilde\rho_{n_1n_2,n'_1n'_2}=\langle n_1n_2|\rho|n'_1n'_2\rangle$ governed by Eq. (\ref{LindbladEff}). Then 
%\begin{eqnarray}\label{IJosephson}
%I(t)&=&\frac{4e}{\hbar}\sum_{n_1,n_2=\{0,1\}}\tilde\rho_{n_1n_2,n_1n_2}(t)~\partial_\phi E_{n_1n_2}(\phi(t)),
%\end{eqnarray}
%where the many body energies $E_{n_1,n_2}$ are the Majorana branches of Fig. \ref{fig:1}b.
%Solving $\tilde\rho(t)$ with Eq. (\ref{LindbladEff}), we compute the Josephson current under bias V, assuming the junction is initially in its ground state $|\Omega(0)\rangle$ (at time $t=0$). 
The goal is to identify smoking-gun features in $I(t)$, in the form of fractional frequency components, that may unambiguously determine the existence of MBS in the TS phase. It has been shown, however, that in the presence of finite dissipation (e.g. non-zero parity mixing), the current becomes, in the long time limit, strictly $2\pi$-periodic, as a consequence of Floquet theorem, and that as a result nothing remains in the stationary current to qualitatively distinguish the trivial from the TS phase \cite{Badiane:PRL11}. 
The two phases, however, exhibit crucial differences in their spectra that give rise to very different features in their \emph{transient} regimes, $I(t<t_T)$. 
%(Transient time $t_T$ is the typical decay time into the quasicontinuum.)
%The differences arise as a result of the sharp anticrossing at $\phi=\pi$, and the finite detachment in the TS phase between the \emph{two} lower ABS and the quasi-continuum. In the trivial phase, in contrast, at least one of the two ABS merges into the quasi-continuum at $\phi=0$ (modulo $2\pi$, see orange curve in Fig. \ref{fig:2}a). 
The spectral features of the TS phase (sharp $\phi=\pi$ anticrossing and detachment of the Majorana sector from the quasicontinuum in the TS phase, see Fig. \ref{fig:2}) allows the even and odd Majorana subspaces in the TS phase to evolve almost coherently performing $4\pi$-periodic population oscillations during a potentially long transient time $t_T\gg T_{2\pi}$, where $T_{2\pi}=2\pi/\omega_J$ is the period of the Josephson driving. We now characterize the transient regime in detail.
%This spectral difference leads to distinct system evolutions, although only in the transient regime (denoted by blue and red colours in Fig. \ref{fig:3}). 
%Indeed, by virtue of the dissipation introduced by the presence of the quasi-continuum, the long time current converges to a stationary $2\pi$-periodic dependence (green coloured plots in Fig. \ref{fig:3}) as a consequence of the Floquet theorem. At this point, there is nothing in the stationary current to distinguish both phases\cite{Badiane11,Yuli}. 
%
%The transient regime that precedes the stationary solution is characterised by a time $t_T$ that can be extremely long for low voltages, $t_T\gg T_{2\pi}$ (see next section). In the TS phase, such transient or \emph{coherent} regime, in which the low energy even and odd Majorana spinors of Fig. \ref{fig:1}b evolve almost coherently without being affected by the environment, may exhibit either of two different behaviors. 
%The evolution of the TS junction depends strongly on the bias voltage. 
At bias $2eV<\delta_\pi$, the junction will remain in the initial ground state $|00\rangle$, resulting in an adiabatic $2\pi$-periodic current (blue curve in Fig. \ref{fig:3}c). % arising from $\partial_\phi E_{00}(\phi)$. 
Increasing $2eV$ above $\delta_\pi$ induces LZ transitions at $\phi=\pi$ from $|\downarrow_e\rangle=|00\rangle$ into $|\uparrow_e\rangle=|11\rangle$, which tend to produce a perfect population inversion as $2eV\gg\delta_\pi$. Then, by virtue of the finite detachment $\delta_0$, if $V$ remains smaller than a certain $V_{\rm{MAR}}$ (quantified below), the population escape into the quasicontinuum as $\phi$ crosses $2\pi$ will be negligible, and the junction will subsequently re-invert into its ground state $|\downarrow_e\rangle$ at $\phi=3\pi$. The repetition of this inversion process gives rise to a transient $4\pi$-periodic Josephson current $I\sim\sin\phi/2$ (red curve in Fig. \ref{fig:3}c), with a superimposed beating envelope (due to imperfect LZ transitions at $\pi$, $3\pi$, etc.). The parameter window required by this solution, $\delta_\pi\ll 2eV\ll 2eV_{\rm{MAR}}$, is experimentally relevant, since $\delta_\pi$ and $V_{\rm{MAR}}$ may be independently controlled by the length $L_{S'}$ and e.g. the junction transparency, respectively.
%%%%%%%%%%%%%%%%%%%%%%%%%%%%%%%%%%%%%%%%%%%%%%%%%%
\begin{figure}
  \centering
  \includegraphics[width=0.48\textwidth]{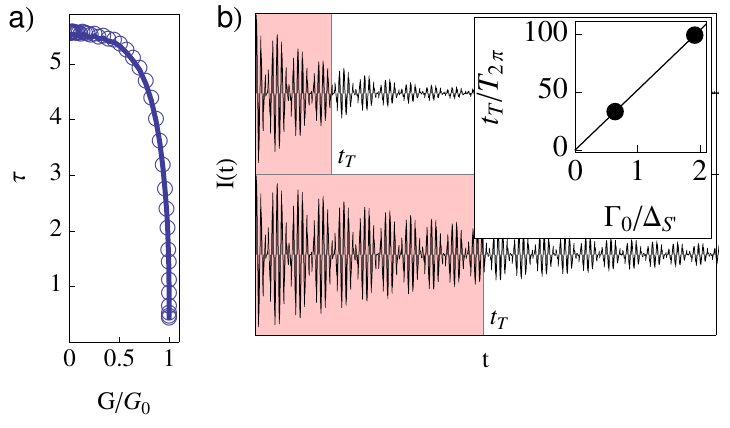}
 \caption{Transient time scales as $t_T\equiv\frac{\hbar^2 \Gamma_0}{(2eV)^2}\tau$, with dimensionless function $\tau(G/G_0)$ plotted in (a). The transient $4\pi$ regime (red shaded region, b) becomes longer as the relaxation rates to the quasicontinuum $\Gamma_0$ increase (Quantum Zeno effect). The two main panels in (b) show the ac Josephson current vs. time ($2eV=0.03\Delta_{S'}$) and two values of $\Gamma_0$ highlighted in the inset, which shows the transient time $t_T\propto \Gamma_0$.}
\label{fig:4}
\end{figure}
%%%%%%%%%%%%%%%%%%%%%%%%%%%%%%%%%%%%%%%%%%%%%%%%%%
In contrast, in the topologically trivial phase, only the lowest of the two Andreev levels is detached from the quasicontinuum. Thus, the state $|11\rangle$ will immediately decay into $|10\rangle$ upon crossing $\phi=2\pi$. As a consequence, the $4\pi$-periodic current cannot develop. %\footnote{Note, however, that any physical mechanism suppressing the escape into the quasi-continuum, e.g. charging effects in a Cooper pair box setup, may restore a $4\pi$ signal even in the trivial phase.} 
The anomalous $4\pi$-periodic component in the transient regime, therefore, is an unequivocal signature of the existence of MBS in the junction. 
At high $V\sim V_{\rm{MAR}}$, the transition into the continuum becomes significant in each period $T_{2\pi}$, for both phases. For such voltages, $t_T\sim T_{2\pi}$, and close to two fermions escape into the contacts per cycle. This process yields a finite dc current (green curve in Fig. \ref{fig:3}c), which is the analogue of the multiple Andreev reflection (MAR) mechanism of conventional junctions. It should be noted that the current for \emph{any} finite bias $V$ eventually develops, in the stationary regime, a finite dc component, that has the same origin as MAR, i.e., the promotion of Cooper pairs into the quasicontinuum by iterated scattering processes (see Fig. \ref{fig:3}e).

A parametric diagram of the MAR (green), the $2\pi$-adiabatic (blue), and the $4\pi$-beating (red) regimes possible are presented in Fig. \ref{fig:3}a,b. The boundaries between regimes are crossovers, defined by $V=V_{\rm{MAR}}$ and $2eV=\delta_\pi$. The different regimes may be clearly distinguished using a finite time spectral analysis of the current, shown in Fig. \ref{fig:3}d. The adiabatic regime exhibits components at $\omega=n\omega_J$, for integer $n>0$, while the MAR regime  develops an additional $\omega=0$ peak. The $4\pi$-periodic transient regime in the TS phase, on the other hand, exhibits a distinct half-integer spectrum $\omega=(n+\frac{1}{2})\omega_J$, together with a peak splitting due to the accompanying beating envelope. The splitting $\pm\delta\omega$ is plotted in Fig. \ref{fig:3}d (inset) as a function of the applied bias, which reveals oscillations due to the interference of successive LZ processes at odd multiples of $\phi=\pi$. 
%The evolution between the transient into the stationary regimes is shown in Fig. \ref{fig:3}d. The typical transient time $t_T\sim 1/\Gamma$ is shown to diverge as $1/V^2$ in Fig. \ref{fig:3}e (note that $T_{2\pi}\propto 1/V$), in agreement with the $\omega_J^2$ dependence of $\Gamma$ in Eq. (\ref{Gamma}).
\paragraph{Transient times.--}% Zeno effect and parity fixing}
The characteristic decay time of the transient current $I(t)\sim e^{-t/t_T}$ is given by %Since this exponential decay is produced by the decay matrix $\Gamma$ in Eq. (\ref{Gamma}), 
%$t_T$ scales with bias $V$ and characteristic relaxation rate $\Gamma_0$ as (\cite{SI, condmatPSJosephson effecttal} information)
%\begin{equation}\label{TT}
%t_T\equiv\frac{\hbar^2 \Gamma_0}{(2eV)^2}\tau, 
%\end{equation}
$t_T\equiv\frac{\hbar^2 \Gamma_0}{(2eV)^2}\tau$, which scales with bias $V$ and the characteristic relaxation rate $\Gamma_0$ that models the escape of quasiparticles from the quasicontinuum levels into the reservoirs \cite{SI}. 
The dimensionless function  $\tau(G/G_0)$, that depends solely on the junction's normal conductance, is numerically computed in Fig. \ref{fig:4}a. It is maximum in the tunneling regime and it decays to zero as transparency goes to one. This means that $t_T$ decreases when the detachment $\delta_0\rightarrow 0$, since then, the escape probability into the quasicontinuum in each cycle goes to one.
%Such result is in accordance with the naive estimate of $p$ in terms of the LZ probability $p=e^{-\pi\delta_0/eV}$. For non-perfect transparency, however, such LZ analysis would predict a characteristic escape voltage $2eV_{\rm{MAR}}=\delta_0$ (red dashed line in Fig. \ref{fig:3}b), which can be quite smaller than the correct result (solid red line), 
The characteristic escape voltage $V_{\rm{MAR}}$ is determined from the condition $t_T=T_{2\pi}$, $2eV_{\rm{MAR}}=\hbar\Gamma_0\tau/2\pi$. One can see that, while a sufficiently slow driving $V$ will suppress quasiparticle poisoning and will thus increase the duration of the transient regime, a complementary, and potentially preferable route is to engineer the environment's $\Gamma_0$ in order to increase the lifetime of the Majorana qubits without slowing operation. Indeed, by increasing the escape rates, parity mixing will be suppressed, increasing $t_T$. This is known as the \emph{quantum Zeno effect}, whereby a quantum system coupled to a bath may develop longer coherence times when increasing the speed at which information is lost into the bath. In a very fast environment, the transitions into the quasicontinuum at $\phi=2\pi n$ will be suppressed by this same Zeno mechanism. This  is demonstrated in the simulation of Fig. \ref{fig:4}b, which shows how $t_T$ is tripled as a result of increasing the $\Gamma_0$. Using realistic parameters for InSb nanowires, we estimate that typical $t_T$ reach into the $\mu$s range at $\mu$V bias voltages. 

Owing to these long transients, the spectrum of microwave radiation should show clear features of the fractional frequencies \cite{SI}.  Such measurement can be performed with an on-chip detector by using, for example, the photon assisted tunneling current of quasiparticles across a superconductor-insulator-superconductor junction capacitively coupled to the TS one. Importantly, it has been \emph{already} demonstrated \cite{Billangeon:PRL07} that such technique allows a direct detection of fractional Josephson frequencies which paves the way for MBS on-chip detection \cite{SI}.
%
%A rough estimate relating the quasiparticle  escape velocity to the group velocity of the normal phase yields $\Gamma_0\approx \Delta_{\rm{eff}}$, which suggests that increasing $\Delta_{\rm{eff}}$ will enable Zeno-type parity protection of driven Majorana qubits. Typical transient times for InSb are then expected to reach into the $\mu$s range at $\mu$V bias voltages.

 %TC:ignore

%\acknowledgments
%\sect{Acknowledgments.}
We are grateful to Y. V. Nazarov, S. Frolov, L. Kouwenhoven and S. Kohler for useful discussions. We acknowledge the support from the CSIC JAE-Doc program and the Spanish
Ministry of Science and Innovation through grants FIS2008-00124, FIS2009-08744.

\bibliography{biblio} 

\begin{thebibliography}{36}
\bibitem [{res()}]{resonantJosephson}%
  \BibitemOpen
  \href@noop {} {}\bibinfo {note} {Note that a close analogy exists between
  Majorana and zero-energy ABS, although the latter are unconstrained by
  fermion parity \cite{SI}.}\BibitemShut {Stop}%
\bibitem [{Maj()}]{MajoranaQubits}%
  \BibitemOpen
  \href@noop {} {}\bibinfo {note} {These parity states constitute an Ising
  qubit, $\sigma_z=-2i\gamma_1\gamma_2$. One can represent a full spin with
  three MBS \cite{Tsvelik:PRL92,Shnirman:PRL03} and thus a logical qubit may be
  constructed \cite{Nayak:RMP08}: $\sigma_z=-2i\gamma_1\gamma_2;
  \sigma_y=-2i\gamma_3\gamma_1; \sigma_x=-2i\gamma_2\gamma_3$.}\BibitemShut
  {Stop}%
\end{thebibliography}
\clearpage


\section*{Supplementary material}

# Supplemental Information for "ac Josephson effect in finite-length nanowire junctions with Majorana modes"

Pablo San-Jose[1], Elsa Prada[2], Ramón Aguado[2]

[1]*Instituto de Estructura de la Materia (IEM-CSIC), Serrano 123, 28006 Madrid, Spain*
[2]*Instituto de Ciencia de Materiales de Madrid (ICMM-CSIC), Cantoblanco, 28049 Madrid, Spain*
(Dated: June 19, 2012)

## ANALOGY WITH THE JOSEPHSON EFFECT THROUGH A SINGLE RESONANT LEVEL MODEL

The bound states of a junction formed by a resonant level, of width $\Gamma$ and energy $\varepsilon$, sandwiched between two superconductors, are the solutions of [1]

$$[\omega - \varepsilon + \frac{\Gamma\omega}{\sqrt{\Delta^2-\omega^2}}][\omega + \varepsilon + \frac{\Gamma\omega}{\sqrt{\Delta^2-\omega^2}}]$$
$$-\Gamma^2 \cos^2\frac{\phi}{2}\frac{\Delta^2}{(\Delta^2-\omega^2)} = 0, \quad (1)$$

which in the resonant case, $\varepsilon = 0$, reduces to

$$\omega \pm \Delta \cos\frac{\phi}{2} + \frac{\omega\sqrt{\Delta^2-\omega^2}}{\Gamma} = 0. \quad (2)$$

The corresponding eigenstates are not single fermions but rather electron-hole superpositions, namely Andreev bound states. The solutions of Eq. (2) become particularly simple in the limit $\Gamma \gg \Delta$ where the Andreev levels can be written as

$$\omega_A \simeq \pm\Delta[1 - 2(\frac{\Delta}{\Gamma})^2]\cos\frac{\phi}{2}. \quad (3)$$

For $\varepsilon \neq 0$, the Andreev bound states can still be written analytically in the limit $\frac{\Delta}{\Gamma} \to 0$ as:

$$\omega_A \simeq \pm\Delta\sqrt{1 - \tau\sin^2\frac{\phi}{2}}, \quad (4)$$

with transmission $\tau = \frac{1}{1+(\frac{\varepsilon}{\Gamma})^2}$.

In the general case, the Andreev levels of this problem exhibit anti-crossings at $\phi = \pi$ for $\tau \neq 1$ (namely $\varepsilon \neq 0$) and a finite detachment from the gap edges at $\phi = 2\pi n$ as $\frac{\Delta}{\Gamma}$ changes. Interestingly, as one approaches $\varepsilon \to 0$, the junction becomes transparent $\tau \to 1$ and one gets Andreev levels that *cannot* be distinguished from the ones of a Majorana junction like the one described in the main text. The deep difference between both systems appears when one takes into account fermion parity conservation when calculating the Josephson current. In the trivial resonant level model, the crossing at $\phi = \pi$ is not protected. Since the bound states have not definite fermion number, transitions between positive and negative energy solutions are allowed. Thus the system evolves across $\phi = \pi$ by remaining in the ground state. As a result, the Josephson current is $2\pi$-periodic. In the Majorana case, the zero-energy bound state is a *true* fermion resonance formed by fusing two Majoranas (each of them non-locally entangled with the corresponding one at the nanowire end) such that the constraint of fermion parity conservation leads to a protected crossing at $\phi = \pi$ and hence the system remains in an excite state after the crossing. This leads to a $4\pi$-periodic current in the ideal infinite wire case, as discussed in the main text. We emphasize again that this deep difference between both systems originates from parity conservation and hence cannot be established by level spectroscopy alone.

Similar considerations apply to junctions in which a Kondo resonance lies between two superconductors. In such case, the Kondo resonance is always pinned at $\varepsilon = 0$ and, again, the Andreev level spectrum shows both crossings at $\phi = \pi$ and detachment from the gap [2, 3].

## FULL ANDREEV LEVEL SPECTRUM FROM BOGOLIUBOV DE GENNES EQUATIONS

The computation of the Andreev levels in an SS'NS'S junction is performed within the Nambu formulation of the junction Hamiltonian

$$H = \frac{1}{2}\int dx \left[\psi_\sigma^+(x), \psi_\sigma(x)\right] \quad (5)$$
$$\times \begin{bmatrix} H^{(0)}_{\sigma\sigma'}(x) & \Delta_{\sigma\sigma'}(x) \\ -\Delta^*_{\sigma\sigma'}(x) & -H^{(0)}_{\sigma\sigma'}(x)^* \end{bmatrix} \begin{bmatrix} \psi_{\sigma'}(x) \\ \psi^+_{\sigma'}(x) \end{bmatrix},$$

where matrix $H^{(0)}(x)$ models the one-dimensional semiconducting wire plus left and right leads, without superconducting pairing. It includes a constant transverse spin-orbit coupling $\alpha$, a Zeeman field along the wire $\mathcal{B} = g\mu_B B/2$ (where $B$ is an external magnetic field, $\mu_B$ is the Bohr magneton and $g$ is the nanowire g-factor), and a position-dependent band shift $\mu(x)$ that accounts for a larger electronic density in the leads than in the wire[4, 5],

$$H^{(0)}_{\sigma\sigma'}(x) = \frac{-\partial_x^2}{2m^*} + i\alpha\sigma_y\partial_x + \mathcal{B}\sigma_x - \mu(x).$$

We approximate $\mu(x)$ by a piecewise constant function, $\mu_{S'}$ in the wire, $\mu_N$ in the central normal region (controllable via a gate voltage $V_g$) and $\mu_S$ in the leads, see Fig. 1a in the main text. The superconducting pairing also

varies with $x$,

$$\Delta_{\sigma\sigma'}(x) = -i\Delta(x)\sigma_y.$$

$\Delta(x)$ is assumed weaker in the superconducting part of the wire than in the leads $|\Delta_{S'}| < |\Delta_S|$, and zero in the central $N$ region. We also denote the phase difference across the $N$ region by $\phi$, so that $\arg[\Delta(\pm|x|)] = \pm\phi/2$.

To compute the Andreev spectrum as a function of $\phi$ one may discretize the integral in Eq. (5), which approximates the continuum problem by a Nambu tight-binding chain. Exact diagonalization of the corresponding $H$ matrix yields $H = \frac{1}{2}\sum_n (d_n^+ d_n - d_n d_n^+)\varepsilon_n$, where the single particle spectrum $\varepsilon_n(\phi)$ are the energies of the Andreev states $|n(\phi)\rangle = d_n^+(\phi)|\Omega(\phi)\rangle$ discussed in the main text, and $|\Omega(\phi)\rangle$ is the ground state. Alternatively, one may employ a wavematching method on the continuum model, and find the energies that yield normalizable solutions in the absence of incoming modes in the leads. We have checked that both methods yield the same results for a sufficiently fine tight-binding discretization, and exhibit band reconnection and Majorana branches if $\mathcal{B} > \sqrt{\Delta_{S'}^2 + \mu_{S'}^2}$, see Fig. 2b in the main text. The effective gap in the topological phase, $\Delta_{\text{eff}}$, is the smallest of the two gaps $\Delta_1 \equiv |\mathcal{B} - \mathcal{B}_c|$ and $\Delta_2 \equiv \Delta_{S'}\sqrt{\frac{2(1+\tilde{\mu}+\sqrt{1+2\tilde{\mu}+\tilde{\mathcal{B}}^2})}{\tilde{\mathcal{B}}^2+2(1+\tilde{\mu}+\sqrt{1+2\tilde{\mu}+\tilde{\mathcal{B}}^2})}} + \mathcal{O}(\Delta_{S'}^2)$, with tilde quantities denoting energies in units of the SO energy, $E_{so} \equiv \hbar^2/ml_{so}^2$, with $l_{so} \equiv \hbar^2/m^*\alpha$ the SO length given in terms of the wire's effective mass $m^*$ and SO coupling $\alpha$.

## NON-ADIABATIC DYNAMICS

As a result of a bias $V$ across the junction, the phase $\phi$ becomes time dependent, $\phi(t) = 2eVt/\hbar = \omega_J t$, and the system is driven out of equilibrium. For a given phase $\phi$, one may build a set of many body instantaneous eigenstates $\{|\varphi_{\vec{n}}(\phi)\rangle = d^{+n_N} \cdots d^{+n_1}|0\rangle\}$, where $n_m = 0, 1$ is the occupation of the $m$-th Andreev level and $H(\phi)|\varphi_{\vec{n}}(\phi)\rangle = E_{\vec{n}}(\phi)|\varphi_{\vec{n}}(\phi)\rangle$. The many body instantaneous energy is $E_{\vec{n}}(\phi) = \sum_m \varepsilon_m(\phi)(n_m - 1/2)$.

The time evolution of the system's full density matrix $\hat{\rho}(t)$ may be expressed in the $\{|\varphi_{\vec{n}}(t)\rangle\}$ basis, $\rho_{\vec{n}\vec{n}'}(t) \equiv \langle\varphi_{\vec{n}}(t)|\hat{\rho}(t)|\varphi_{\vec{n}'}(t)\rangle$, and is governed by the equation of motion,

$$\partial_t \rho(t) = -\frac{i}{\hbar}[\mathcal{H}(t), \rho(t)],$$

The effective Hamiltonian $\mathcal{H}(t)$ includes the connection between instantaneous eigenstates, $\mathcal{A}_{\vec{n}'\vec{n}}(\phi) \equiv i\langle\varphi_{\vec{n}'}(\phi)|\partial_\phi \varphi_{\vec{n}}(\phi)\rangle$,

$$\mathcal{H}_{\vec{n}\vec{n}'}(t) = \delta_{\vec{n},\vec{n}'} E_{\vec{n}}(\phi(t)) - \hbar[\partial_t \phi(t)]\mathcal{A}_{\vec{n}\vec{n}'}(\phi(t)).$$

The many body connection $\mathcal{A}_{\vec{n}\vec{n}'}$ may be related to the single particle connection $A_{nm} \equiv i\langle n|\partial_\phi|m\rangle = i\langle\Omega|d_n(\partial_\phi d_m^+)|\Omega\rangle$ [6] and the anomalous connection $\bar{A}_{nm} \equiv i\langle\Omega|(\partial_\phi d_m^+)d_n^+|\Omega\rangle$ by

$$i\langle\varphi_{\vec{n}}|\partial_\phi|\varphi_{\vec{n}}\rangle = \sum_m A_{n_m n_m} \quad (6)$$

$$i\langle\varphi_{\vec{n}}|\partial_\phi d_i^+ d_j|\varphi_{\vec{n}}\rangle = -i\langle\varphi_{\vec{n}}|\partial_\phi d_i d_j^+|\varphi_{\vec{n}}\rangle^* = A_{ji} \quad (7)$$

$$i\langle\varphi_{\vec{n}}|\partial_\phi d_i^+ d_j^+|\varphi_{\vec{n}}\rangle = -i\langle\varphi_{\vec{n}}|\partial_\phi d_i d_j|\varphi_{\vec{n}}\rangle^* = \bar{A}_{ji} \quad (8)$$

for any state $|\varphi_{\vec{n}}\rangle$. All other connections are zero, which implies that the system evolution preserves fermion parity (i.e. if the initial number of excitations is definitely even or odd, it will remain that way, even though the total number may change due to the anomalous connection $\bar{A}$.)

Driving may impart the excitations with energy, which may coherently promote them into the quasi-continuum above $\Delta_{\text{eff}}$. To account for such quasiparticle poisoning, one could pursue a brute-force approach by including an extended basis with all many-body states generated by filling the first $N$ Andreev levels in the spectrum. [7] This is impractical, however, since the size of the basis grows exponentially as $2^N$. Moreover, it soon becomes clear that once a fermion fully escapes from the Majorana sector, it has a vanishing probability of returning, given the large phase space available in the quasi-continuum. Thus, these excitations are irreversible to all practical effects. This single-fermion escape process changes parity. To model such two-step quasiparticle *poisoning* process without having to track the dynamics of the (exponentially) large number of many body states, one may substitute the single-particle quasi-continuum ($h$, or 'high') levels $|m\rangle_h$ by dissipative levels (with zero mutual connection) that decay directly into a fermion reservoir at a rate $\Gamma_0^{(m)}$, while the 'low' levels ($l$, below the gap) remain non-dissipative. If we assume a Markovian approximation for the decay of $h$ into the fermionic bath, the master equation takes on a Lindblad form

$$\partial_t \rho = -\frac{i}{\hbar}[\mathcal{H}, \rho] + \sum_m \Gamma_0^{(m)}\left(L_m \rho L_m^+ - \frac{1}{2}\{L_m^+ L_m, \rho\}\right)$$

If we further constrain our dynamical space to states with a total of 1 or 0 fermions in the $h$ sector, $\{|\varphi_{\vec{n}_l;1_m}\rangle, |\varphi_{\vec{n}_l;0}\rangle\}$ (fast decay limit), the Lindblad operators $L_m$ will simply project any state with a fermion in the $m$-th dissipative level ($m \in h$), into another without it $L_m = \sum_{\vec{n}_l}|\varphi_{\vec{n}_l;0}\rangle\langle\varphi_{\vec{n}_L;1_m}|$.

For large enough $\Gamma_0^{(m)} \sim \mathcal{O}(\eta^{-1})$ ($\eta$ being a small perturbative parameter), the irreversible decay will suppress the density matrix in the quasi-continuum, $\rho_{ll} \sim \mathcal{O}(1)$, $\rho_{lh}, \rho_{hl} \sim \mathcal{O}(\eta)$, $\rho_{hh} \sim \mathcal{O}(\eta^2)$. One may then perturbatively solve the detailed balance conditions for the $h$ sector, which eventually leads to a Lindblad-type master

equation for the reduced density matrix $\rho_{ll}$,

$$\partial_t \rho_{ll} = -\frac{i}{\hbar}[\mathcal{H}_{ll}, \rho_{ll}]$$
$$+ \sum_{\alpha\beta} \Gamma_{\alpha\beta}\left(\mathcal{L}_\alpha \rho_{ll}\mathcal{L}_\beta^+ - \frac{1}{2}\left\{\mathcal{L}_\beta^+\mathcal{L}_\alpha, \rho_{ll}\right\}\right) + \mathcal{O}(\eta)$$

Considering the $l$ sector as spanned by two Andreev levels, of energies $\varepsilon_{1,2}$, the corresponding many body basis $|\varphi_{n_1,n_2;0}\rangle \equiv |n_1 n_2\rangle$ is $\{|\downarrow_e\rangle, |\uparrow_e\rangle, |\downarrow_o\rangle, |\uparrow_o\rangle\} \equiv \{|00\rangle, |11\rangle, |10\rangle, |01\rangle\}$, where $e$ and $o$ stand for even and odd fermion parity. Then, the Lindblad equation above reduces to

$$\partial_t \tilde{\rho} \approx -\frac{i}{\hbar}[\mathcal{H}, \tilde{\rho}] + \sum_{\alpha\beta} \Gamma_{\alpha\beta}\left(\mathcal{L}_\alpha \tilde{\rho}\mathcal{L}_\beta^+ - \frac{1}{2}\left\{\mathcal{L}_\beta^+\mathcal{L}_\alpha, \tilde{\rho}\right\}\right) \quad (9)$$

with four parity-mixing Lindblad operators,

$$\begin{aligned}
\mathcal{L}_1 &= |\downarrow_e\rangle\langle\uparrow_o| + |\downarrow_o\rangle\langle\uparrow_e| \\
\mathcal{L}_2 &= |\downarrow_e\rangle\langle\downarrow_o| - |\uparrow_o\rangle\langle\uparrow_e| \\
\mathcal{L}_3 &= -|\uparrow_e\rangle\langle\uparrow_o| + |\downarrow_o\rangle\langle\downarrow_e| \\
\mathcal{L}_4 &= |\uparrow_e\rangle\langle\downarrow_o| + |\uparrow_o\rangle\langle\downarrow_e|
\end{aligned} \quad (10)$$

and a $4\times 4$ relaxation matrix

$$\Gamma_{\alpha\beta}(\phi) = 4\omega_J^2 \sum_m \frac{\nu_{m\alpha}^*(\phi)\nu_{m\beta}(\phi)}{\Gamma_0^{(m)}}. \quad (11)$$

The $\vec{\nu}_m(\phi)$, given explicitly in terms of single-particle connections between the $l$ states and the $h$ states,

$$\vec{\nu}_n = \left\{A_{2,n}, A_{1,n}, \bar{A}_{1,n}, \bar{A}_{2,n}\right\}, \quad (12)$$

are peaked at around $\phi = 2\pi n$, for integer $n$. They quantitatively account for parity mixing mediated by coherent excitation into dissipative level $m$, and contain detailed microscopic information about the quasi-continuum, making the resulting dynamics for the reduced density matrix highly non-trivial.

Since the levels $\varepsilon_{m\geq 3}$ above $\Delta_\text{eff}$ are almost $\phi$ independent, the Josephson current through the biased junction may be approximated in terms of the $4\times 4$ reduced density matrix in the Majorana sector, namely the $\tilde{\rho}_{n_1 n_2, n_1' n_2'} = \langle n_1 n_2 |\rho| n_1' n_2'\rangle$ governed by Eq. (9). Then

$$I(t) = \frac{4e}{\hbar} \sum_{n_1,n_2=\{0,1\}} \tilde{\rho}_{n_1 n_2, n_1 n_2}(t)\ \partial_\phi E_{n_1 n_2}(\phi(t)), \quad (13)$$

where the many body energies $E_{n_1,n_2}$ are the Majorana branches of Fig. 1b in the main text. Solving $\tilde{\rho}(t)$ with Eq. (9), we compute the Josephson current under bias V, assuming the junction is initially in its ground state $|\Omega(0)\rangle$ (at time $t=0$).

## DISCUSSION ABOUT EXPERIMENTAL DETECTION

Parity protection by the Zeno effect should prove useful for prolonging the $4\pi$-periodic transient regime when the main source of quasiparticle poisoning is through the quasi-continuum. Other sources of poisoning not discussed here include non-equilibrium quasiparticles from the rest of the circuit, a contrition that can, however, be greatly reduced by including quasiparticle traps, such as nearby metal contacts [8]. In the case of quasicontinuum poisoning, a rough estimate relating the quasiparticle escape velocity to the Fermi velocity yields $\Gamma_0 \approx \Delta_\text{eff}$, which suggests that increasing $\Delta_\text{eff}$ will improve Zeno-type parity protection of driven Majorana qubits. Using realistic parameters for InSb nanowires, we estimate that typical transient times reach into the $\mu$s range at $\mu$V bias voltages.

The spectrum of microwave radiation from TS junctions should show clear features of the fractional frequencies (Fig. 3d, main text). Such measurement can be performed with an on-chip detector, which greatly minimizes the impedance-matching problems in classical detection schemes. This on-chip detection can be achieved by using, for example, the photon assisted tunneling current of quasiparticles across a superconductor-insulator-superconductor junction capacitively coupled to the TS one. It has been *already* demonstrated [9] that such technique allows a direct detection of fractional Josephson frequencies [10] of a superconducting single electron transistor [11]. This is possible if $T_1 \gg t_T$, where $T_1$ is the relaxation time corresponding to the transition $|\uparrow_e\rangle = |11\rangle \to |\downarrow_e\rangle = |00\rangle$ that brings the system back to the ground state. Such parity-conserving transitions, that at low enough temperatures are due to quantum fluctuations of the phase, can be attributed to photon emission to the electromagnetic environment and, to a lesser amount, to phonon emission. Proper engineering of the circuit containing the topological wire can, in principle, greatly reduce both emission processes. For example, photon emission can be inhibited by reducing the ohmic component of the impedance seen by the Majorana qubit [12]. To lowest order, this relaxation rate reads:

$$\begin{aligned}
T_1^{-1} &= \frac{2\hbar}{\Delta E_e}\frac{1}{4e^2}|\langle\uparrow_e |\hat{I}| \downarrow_e\rangle|^2 \frac{\text{Re}\left[Z(\Delta E_e/\hbar)\right]}{R_K} \quad (14)\\
&= \frac{8\Delta E_e}{\hbar}|\langle\uparrow_e |\partial_\phi| \downarrow_e\rangle|^2 \frac{\text{Re}\left[Z(\Delta E_e/\hbar)\right]}{R_K},
\end{aligned}$$

where $\hat{I} = (4e/\hbar)\partial_\phi\mathcal{H}$, $\Delta E_e \equiv E_{\uparrow_e} - E_{\downarrow_e}$ and $R_K = h/e^2 \approx 25.81 k\Omega$ is the resistance quantum. This rate is $\phi$-dependent. Substituting the value of the connection between the even Majorana branches and their energy difference for the wire considered in Fig. 3 of the main text, and averaging over the phase, we obtain relaxation times of the order of $1\mu$s for impedances of $\text{Re}\left[Z(\Delta E_e/\hbar)\right] = 1\Omega$.



\end{document}